\pdfoutput=1
\documentclass[useAMS,usenatbib]{mn2e}
\usepackage[a4paper,centering, totalwidth=520pt, totalheight=700pt]{geometry}
\usepackage{graphicx}
\usepackage{aas_macros}     
\usepackage{amsmath}
\usepackage{amssymb}
\usepackage{calligra}
\usepackage{bm}
\usepackage{epstopdf}
\usepackage{url}
\usepackage{hyperref}
\usepackage{slantsc}
\hypersetup{
    colorlinks=true,
    linkcolor=black,
    citecolor=black,
}

\newcommand{\tcr}[1]{{#1}}

\title[Can we distinguish early dark energy from a cosmological constant?]
{Can we distinguish early dark energy from a cosmological constant?}

\author[D. Shi and C. M. Baugh]{
\parbox[h]{\textwidth}
{Difu Shi$^{1} \thanks{difu.shi@durham.ac.uk}$,
Carlton M. Baugh$^{1}$.}
\vspace*{6pt} \\
$^1$ Institute for Computational Cosmology, Department of Physics, Durham University, South Road, Durham DH1 3LE, UK.}

\begin{document}

\label{firstpage}

\maketitle

\date{\today}
\pagerange{\pageref{firstpage}--\pageref{lastpage}} \pubyear{2015}

\begin{abstract}
Early dark energy (EDE) models are a class of quintessence dark energy
with a dynamically evolving scalar field which display a small but
non-negligible amount of dark energy at the epoch of matter-radiation
equality. Compared with a cosmological constant, the presence of dark energy at
early times changes the cosmic expansion history and consequently the
shape of the linear theory power spectrum and potentially other observables.
We constrain the cosmological parameters in the EDE cosmology using
recent measurements of the cosmic microwave background and baryon
acoustic oscillations. The best-fitting models favour no EDE; here
we consider extreme examples which are in mild tension with current
observations in order to explore the observational consequences of
a maximally allowed amount of EDE. We study the non-linear evolution
of cosmic structure in EDE cosmologies using large volume N-body
simulations. Many large-scale structure statistics are found to be
very similar between the $\Lambda$ cold dark matter ($\Lambda$CDM)
and EDE models. We find that \tcr{EDE cosmologies predict fewer massive halos in comparison
to $\Lambda$CDM, particularly at high redshifts.} The most promising way to distinguish EDE from
$\Lambda$CDM is to measure the power spectrum on large scales, where
differences of up to 15\% are expected.
\end{abstract}


\section{Introduction}
One of the key objectives of future galaxy surveys is to determine
the nature of the dark energy behind the accelerating cosmic expansion.
In particular, does the dark energy take the form of a cosmological
constant, which is hard to explain from a theoretical perspective,
or is it a dynamical field, with a time dependent equation of state?
What is the best way to distinguish between these scenarios for the
dark energy? Here we demonstrate that this is a remarkably challenging
problem, once the competing models have been set up to reproduce what we
already know about the Universe.

The standard $\Lambda$ Cold Dark Matter ($\Lambda$CDM) cosmological model,
in which dark energy is time independent, provides a good description of
current data \citep[e.g.][]{Efstathiou2002,Sanchez2009,Sanchez2012,
Planck2013XVI,Planck2015XIII}.
However, the cosmological
constant lacks theoretical motivation and throws up issues such as the
fine-tuning and the coincidence problems. Many alternatives have been proposed
to alleviate these problems \citep*[e.g. the review by ][]{Copeland2006}.
A number of these are based on time-evolving scalar fields, which are usually
referred to as quintessence models \citep*{Ratra1988,Wetterich1988,Caldwell1998,
Ferreira1998}.

In $\Lambda$CDM, the impact of the cosmological constant on the cosmic expansion
can be ignored once the energy density of the dark energy falls below $\sim 1\%$
of the critical density, which occurs above $z \sim 5$. In contrast, a class of
quintessence models called early dark energy (EDE) display a small but
non-negligible amount of dark energy at early times which can change the
expansion rate appreciably, even as early as the epoch of matter-radiation
equality. These models can be divided into two classes:
the so called ``tracker fields" \citep*{Steinhardt1999} and ``scaling solutions"
\citep{Halliwell1987,Wetterich1995}.

Previous simulations of EDE cosmologies, such as those by \citet{Grossi2009},
\citet*{Francis2009} and \citet{Fontanot2012}, focused on the impact
on structure formation of the different expansion history with EDE compared
with $\Lambda$CDM, whilst keeping the same linear theory power spectrum
and background cosmological parameters as used in $\Lambda$CDM.
However, to produce a fully self-consistent model two further steps are necessary
in addition to changing the expansion history \citep{Jennings2010}. First,
the best fitting cosmological parameters will be different in EDE cosmologies
than they are in $\Lambda$CDM.
Second, the input power spectrum used to set up the initial
conditions for the N-body simulation should be different in EDE from that
used in $\Lambda$CDM. The change in the expansion history alters the width of
the break in the power spectrum around the scale of the horizon at matter -
radiation equality \citep{Jennings2010}. This change in the power spectrum
is compounded by the changes in the cosmological parameters between the best
fitting EDE and $\Lambda$CDM models. If we are to
compare models that satisfy the current observational constraints to look for
measurable differences which can be probed by new observations, we need to
take all three of these effects into account.

EDE models can be described in terms of a scalar field potential,
with the dynamical properties obtained by minimizing the action that includes the
scalar field potential. We take a more practical view and
consider parametrizations of EDE models which allow us to explore the
parameter space more efficiently. \cite{Corasaniti2003} presented four and
six parameter models for the time dependence of the
equation of state parameter of the dark energy, $w$, which give
very accurate reproductions of the results of the full Lagrangian minimisation.
However, with current data it is not feasible to constrain such a large number of
additional parameters in addition to the standard cosmological parameters.
Instead we investigate two parameter formulations of the dark energy.

We demonstrate that current observations of temperature fluctuations and the
polarization of the cosmic microwave background (CMB) radiation and the
apparent size of baryon acoustic oscillations (BAO) in the galaxy distribution
already put tight constraints on EDE models. In fact, the best fitting models
are consistent with {\it no} early dark energy, a conclusion that has been
reached by other studies \citep{Planck2013XVI,Planck2015XIV}.
Nevertheless, models with appreciable amounts of dark energy remain formally
consistent with the current data. We consider two cases which have one and two
percent of the critical density in dark energy back to the epoch of
matter radiation equality.

In the standard lore, EDE models display a more rapid expansion at high redshift
than $\Lambda$CDM and so, if they are normalised to have the same fluctuations on
$8 h^{-1} {\rm Mpc}$ today (ie the same value of $\sigma_{8}$), structures
form earlier in these models. We find that this is not a generic feature of EDE.
The EDE models we consider have growth rates that are very similar to that in
$\Lambda$CDM, even lagging behind $\Lambda$CDM at intermediate redshifts.
This results in these cosmologies actually displaying {\it fewer} massive
haloes than $\Lambda$CDM at high redshifts.

This paper is organized as follows. In Section~\ref{sec:theory} we discuss the
parametrization of EDE models (\S~2.1), the constraints derived on cosmological
parameters using CMB and BAO data (\S~2.2), compare the rate at which
fluctuations grow in EDE and $\Lambda$CDM (\S~2.3) and describe the N-body
simulations carried out (\S~2.4). The simulation results, namely the matter
power spectrum, distribution function of counts-in-cells and halo mass function
are presented in Section~\ref{sec:result}. Finally, in Section~\ref{sec:conclusion},
we give a summary of our results.

\section{Theoretical background}
\label{sec:theory}

In this section we explain the behaviour of EDE cosmologies and how this
is parametrized (\S~\ref{sec:Model}), and then present constraints on the
cosmological parameters in EDE and $\Lambda$CDM (\S~\ref{sec:parameterfit}).
The rate at which fluctuations grow in the different cosmologies is calculated
in \S~\ref{subsec:GR}. The numerical simulations used are described in
\S~\ref{sec:simulation}.

\subsection{Early Dark Energy cosmologies}
\label{sec:Model}

\begin{figure}
\includegraphics[clip,width=8.5cm]{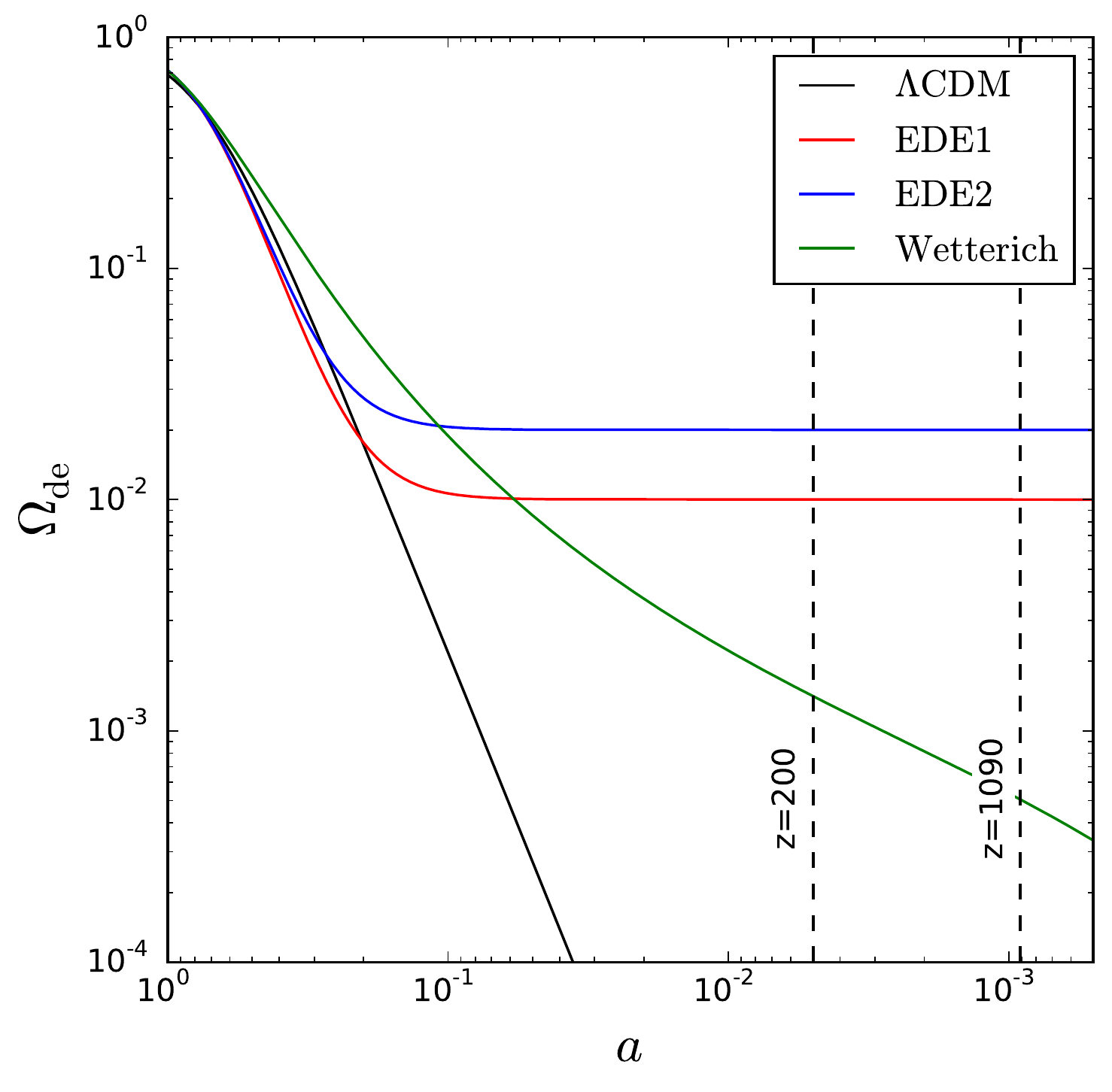}
\caption{The dark energy density parameter, $\Omega_{\rm de}$, as a function of scale factor, $a$, for the two EDE models studied here, the EDE1 model (red line), EDE2 model (blue line), the Wetterich model (green line) and $\Lambda$CDM  (black line). (See Table~1 for the model parameters.) The two black dashed lines indicate, as labelled, redshift 200 when our simulations are started and the CMB redshift, $z\sim$1090.}
\label{fig:Omega_DE}
\end{figure}

The dark energy equation of state, $w(z)=P/\rho$, where $P$ is pressure
and $\rho$ is density, determines how dark energy influences the expansion
of the universe. In the standard $\Lambda$CDM model, the equation of state
of the dark energy is a constant, $w_{\Lambda}=-1$, and the dark energy density
parameter $\Omega_{\rm de}(z)$ falls rapidly to zero with increasing redshift
(see Fig.~\ref{fig:Omega_DE}). The cosmological constant can be completely ignored
beyond {\bf $z \sim 5$}, once it accounts for less than $1\%$ of the
critical density. However, if the dark energy equation of state is such that $w>-1$,
$\Omega_{\rm de}$ will decrease more slowly and the consequences of dark energy
will be felt earlier.

Quintessence originates from theoretical models which treat the dark energy as a
slowly evolving scalar field. The scalar field can be described by potentials with
different properties. Viable models share common features such as reproducing the observed magnitude
of the present-day energy density and producing an accelerating expansion at late times.
Due to the time-dependent scalar field, the dark energy equation of state evolves.
The ratio of the energy density of dark energy to the critical density in quintessence
models, $\Omega_{\rm de}$, will be different from that in the $\Lambda$CDM model.
This affects the growth of structure (see Fig.~\ref{fig:Omega_DE} for a comparison
between $\Omega_{\rm de}$ in $\Lambda$CDM and in the EDE models simulated here;
the choice of EDE model is discussed later in \S~\ref{sec:parameterfit}).
Observations constrain the present-day dark energy equation of state to be
$w_{0}<-0.8$ \citep{Sanchez2012}. So, EDE models which agree with this constraint
should display a transition in $w$ from the present day value ($w \approx -1$) to
the early-time value (usually close to zero). How and when this transition
happens is the main difference between the various EDE models.

Ideally, the dark energy equation of state should be derived from the potential energy
associated with a time dependent scalar field. However, the motivation behind the form
of the potential is weak which means that a wide variety of cases have been considered
\citep{Corasaniti2003}. One way to carry out a systematic study of the EDE parameter
space is to use a parametrization for the dark energy equation of state, $w$, or the
dark energy density parameter, $\Omega_{\rm de}$. This approach offers a
model-independent and efficient way to investigate the properties of EDE models
which display similar behaviour for $w$.

The most commonly used and simplest parametrization to describe the evolution
of the equation of state is the two-parameter equation, $w=w_{0}+(1-a)w_{a}$,
where $a$ is the expansion parameter \citep{Chevallier2001,Linder2003}.
However, \citet*{Bassett2004} have shown that a two-parameter equation is not sufficiently
accurate to describe the equation of state of the scalar field to better than 5 per cent beyond $z \sim 1$.
This problem is even worse when if a two-parameter model is to be used in an
N-body simulation which might start at a very high redshift (e.g. $z \approx 100$).
More complex parametrizations with more parameters have been proposed which can
capture the behaviour seen in a wide range of quintessence models
\citep{Corasaniti2003}. However, the additional parameters are hard to constrain
in practice given current observations.

Instead we investigate empirical parametrizations of EDE which have three
parameters. One was introduced by \citet{Wetterich2004} and is given in terms
of the equation of state parameter,
\begin{equation}
w(a)=-\frac{w_{0}}{(1-b\ln(a))^{2}},
\label{Wetterich}
\end{equation}
where
\begin{equation}
b=-\frac{3w_{0}}{\ln \left(\frac{1-\Omega_{\rm de,e}}{\Omega_{\rm de,e}}\right)+\ln\left(\frac{1-\Omega_{\rm m,0}}{\Omega_{\rm m,0}}\right)}.
\label{Wetterich_b}
\end{equation}
Here $w_{0}$ is the dark energy equation of state today, $\Omega_{m,0}$ is the matter (i.e. baryons and cold dark matter) density parameter at
$z=0$. $\Omega_{\rm de,0}$ and $\Omega_{\rm de,e}$ are, respectively, the dark energy
density parameters today and as $z\to \infty$.

The other empirical parametrization we consider was proposed by \citet{Doran2006} and
is written in terms of the time evolution of the dark energy density parameter
\begin{equation}
\Omega_{\rm de}(a)=\frac{\Omega_{\rm de,0}-\Omega_{\rm de,e}(1-a^{-3w_{0}})}{\Omega_{\rm de,0}+ \Omega_{\rm m,0}a^{3w_{0}}}+\Omega_{\rm de,e}(1-a^{-3w_{0}}).
\label{Doran}
\end{equation}

\begin{figure}
\includegraphics[clip,width=8.5cm]{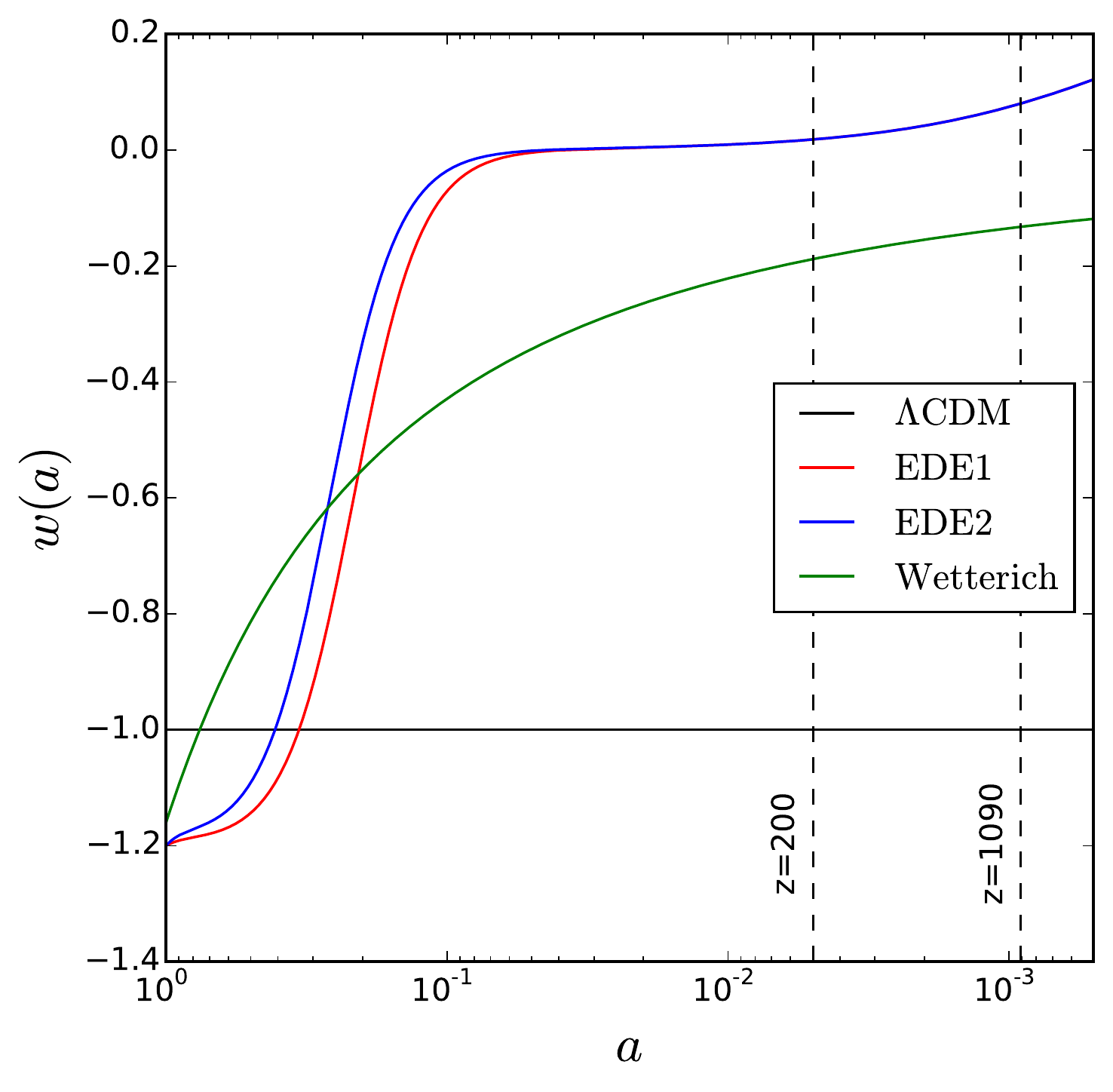}
\caption{The dark energy equation of state, $w$, as a function of the scale factor,
$a$, for the EDE1 model (red line), EDE2 model (blue line), a Wetterich model (green line)
and $\Lambda$CDM  (black line). (See Table~1.) The two black dashed lines indicate the redshift when
the simulations are started ($z = 200$) and the CMB redshift ($z\sim 1090$).}
\label{fig:EoS}
\end{figure}

Both parametrizations mimic $\Lambda$CDM at low redshift and can provide
non-negligible amounts of EDE at early times, depending upon the parameter
values adopted. The Doran \& Robbers parametrization allows rapid transitions
in the dark energy equation of state. The variation of $w(a)$ in the Wetterich
parametrization is more gradual as shown in Fig.~\ref{fig:EoS}. If we assume
$\Omega_{\rm m} + \Omega_{\rm de} =1$ at $z=0$, the two parametrizations yield
$\Lambda$CDM, $w(a)=w_{0}=-1$, in the limit when $\Omega_{\rm de,e}=0$.

\subsection{Parameter fitting}
\label{sec:parameterfit}

\begin{figure}
\includegraphics[width=9cm]{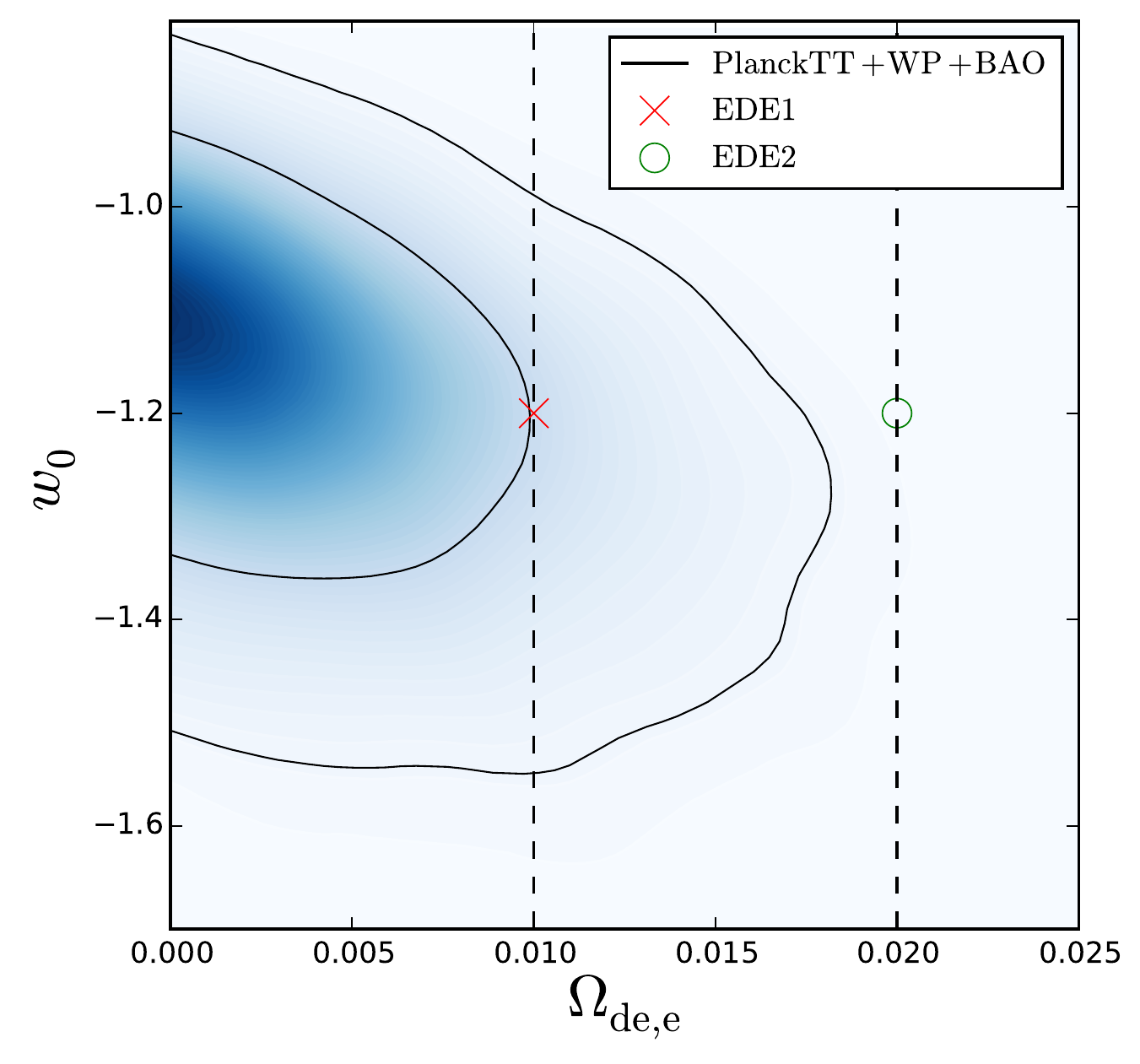}
\caption{The 2D marginalized distribution in the present day equation
of state parameter, $w_{0}$, and the critical density in dark
energy at early times, $\Omega_{\rm de}$, using the Doran \& Robbers EDE
parametrization for the Planck TT, WMAP polarization and BAO data
combination. The constraint is compatible with $\Lambda$CDM. The solid black
lines show the 68\% and 95\% confidence intervals. The red cross and green circle
indicate, respectively, the EDE1 and EDE2 models which are used in our simulations.
The black dashed lines indicate the values of $\Omega_{\rm de}$ in these models.}
\label{fig:w0_Omega}
\end{figure}

Changing the equation of state, $w$, from a constant to being time-dependent will
affect the evolution of the Universe. The cosmological distance-redshift relation
also changes. Cosmological constraints derived for $\Lambda$CDM will not necessarily
apply in an EDE universe. We need to re-fit the cosmological parameters
for an EDE cosmology and use the best-fitting values in a simulation of such a model
rather those derived for $\Lambda$CDM. Here we use observations of the CMB and BAO
to find the best-fitting cosmological parameters for EDE models. Using the CMB and
BAO data in this way not only allows us to determine the cosmological parameters
we should use in simulations, but is also a preliminary test of the viability of
EDE parametrization.

To derive the constraints on EDE parameters, we use the CMB measurement from the
Planck 2013 data release \citep{Planck2013XVI}, which contains the
Planck temperature angular power spectrum (TT) and WMAP9 polarization data (WP),
in the form of likelihood software\footnote{We note that the Planck 2015 results
 show a somewhat tighter constraint on $\Omega_{\rm de,e}$
for the \citet{Doran2006} model than we find using the 2013 data release \citep{Planck2015XIV}.}.
We adapted the Markov Chain Monte-Carlo code, CosmoMC, to work for EDE cosmologies
\citep{CosmoMC}. Some studies, such as \citet{Wang2006}, use CMB
distance priors which condense the full temperature fluctuation power spectrum into three quantities which
depend on an assumed cosmological model to describe the peak positions and peak height
ratios  \citep{Komatsu2009, wang2013}.
Although this method is faster, we do not use it here because it results in weaker
constraints than using the full data set.

We also use the BAO feature in the galaxy distribution which depends on the horizon scale
at matter-radiation decoupling and angular diameter distance to a given redshift. The BAO
measurements used are the $z=0.106$ result from the 6dF Galaxy Survey
\citep[6dFGS,][]{Beutler2011}, the $z=0.35$ measurement from Sloan Digital Sky Survey
Data Release 7 \citep[SDSS DR7,][]{Percival2010} and the $z=0.57$ measurement from
the Baryon Oscillation Spectroscopic Survey \citep[BOSS,][]{Sanchez2012}.

Fig.~\ref{fig:w0_Omega} shows the 2D marginalized distribution for $w_{0}$ and
$\Omega_{\rm de,e}$ using the Doran \& Robbers parametrization of EDE.
$\Lambda$CDM ($w_0=-1$,$\Omega_{\rm de,e}=0$) is within the 68\% confidence level.
The Doran \& Robbers cosmologies with 1 and 2 percent EDE are respectively roughly 1
and 2 $\sigma$ away from the best-fitting value. In order to maximize the
effects of EDE, here, we choose $w_0=-1.2$ and $\Omega_{\rm de,e}=0.01$ as the ``EDE1'' model,
$\Omega_{\rm de,e}=0.02$ as the ``EDE2'' model rather than using the best fitting values
and keep the other cosmological parameters the same between the
two models. The EDE1 and EDE2 models are therefore somewhat in tension with the
current observational constraints but are formally consistent with the data.

Table~\ref{tab:best_fit} summarizes the constraints for the $\Lambda$CDM and
EDE cosmologies, assuming a flat universe. In the EDE models, the cosmological
parameters show small departures from the best fitting $\Lambda$CDM values.
The best fitting result obtained using the Wetterich parametrization gives
a negligible amount of EDE, $\Omega_{\rm de,e}\sim 0$, corresponding to
$\Lambda$CDM if we fix $w_0=-1$. The Wetterich parametrization does not
yield any EDE when constrained using current observations. The Doran \& Robbers
parametrization can reproduce the step-like transition in the dark energy equation
of state that results from solving the equations of motion for an  EDE potential, so we focus
on this parametrization from hereon.

\begin{figure*}
\includegraphics[clip,width=18cm]{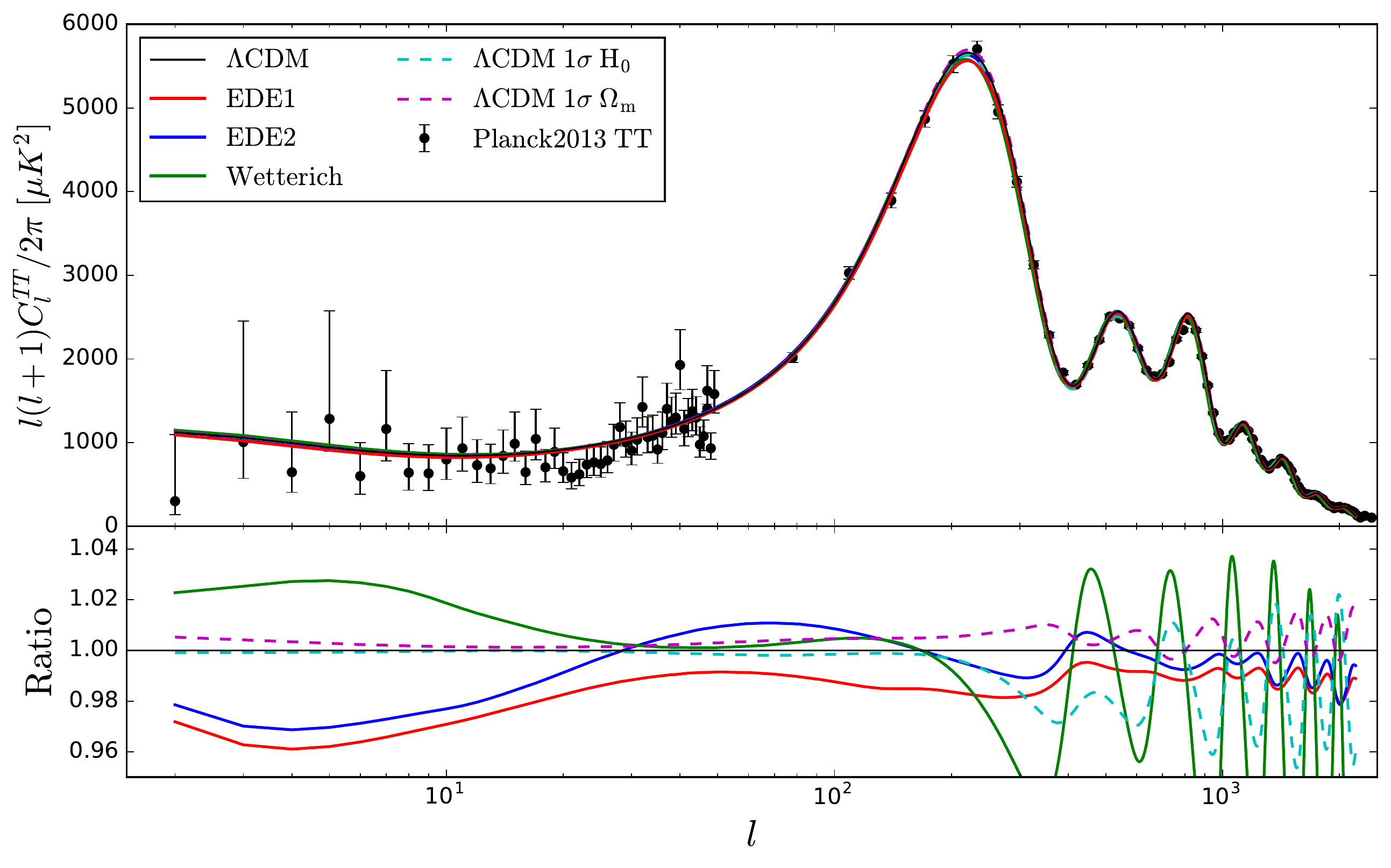}
\caption{The upper panel shows the cosmic microwave background temperature
fluctuation spectra of the EDE1, EDE2, $\Lambda$CDM and Wetterich models from
Table~1 compared with the Planck 2013 data \citep{Planck2013XVI} (see legend).
Two variants of the $\Lambda$CDM model are also shown, which depart from the
best fitting model by similar amounts to the EDE models. The lower panel shows
the ratio of these models to $\Lambda$CDM.
}
\label{CMB}
\end{figure*}

\begin{table}
\begin{center}
\caption{Summary of the best fitting values using CMB and BAO data for the
dark energy parametrizations of Doran \& Robbers (labelled EDE1 and EDE2) and Wetterich, along with
$\Lambda$CDM. All models have $\sigma_{8}=0.8$.}\label{tab:best_fit}
\begin{tabular}{ccccc}
  \hline
  Parameter & $\Lambda$CDM & EDE1 & EDE2 & Wetterich\\
  \hline
  $H_0$ & 67.7 & 71.9 & 71.9 & 70.7\\
  $\Omega_{de}$ & 0.687 & 0.719 & 0.719 & 0.716\\
  $\Omega_b$ & 0.0488 & 0.0424 & 0.0424 & 0.044 \\
  $w_0$  & -1 & -1.2 & -1.2 & -1.16\\
  $\Omega_{de,e}$ & - & 0.01 & 0.02 & $<10^{-7}$ \\
  \hline
\end{tabular}
\end{center}
\end{table}

Fig.~\ref{fig:EoS} shows the dark energy equation of state as a function of
scale factor for the EDE1 and EDE2 models, along with the $\Lambda$CDM model.
The corresponding dark energy density parameter as a function of scale factor
is shown in Fig.~\ref{fig:Omega_DE}.  Here, we plot the Wetterich model
with $\Omega_{\rm de,e}=10^{-5}$, which is much larger than the value listed in
Table~1,  but retain the other best-fitting cosmological
parameters for comparison. At late times the EDE1 and EDE2 models show very
similar behaviour to $\Lambda$CDM, with a rapid transition to $w\approx 0$
at early times. The dark energy parameter remains nearly constant at early
times ($z\gtrsim 9$). Even for the tiny amount of EDE considered, the
Wetterich model deviates from $\Lambda$CDM from very low redshift.
The BAO data probe low redshifts which is why the observational constraints
do not allow Wetterich model to have non-negligible EDE.

Since the EDE1 and EDE2 models are not best-fitting models, in order evaluate
the effect of the deviations before running simulations, we look at two variant
$\Lambda$CDM models for comparison. One is a $\Lambda$CDM model with a value
of $\Omega_{\rm m}$ which deviates by $1\sigma$ from the best-fitting value,
labelled ``$\Lambda$CDM $1\sigma\ \Omega_{\rm m}$''. The other one is a
$\Lambda$CDM model with $H_0$ deviating by $1\sigma$ from the best-fitting value,
named ``$\Lambda$CDM $1\sigma\ H_0$''. We use the {\scshape CAMB} code \citep{Lewis2002}
to generate the CMB temperature spectra for those models. Fig.~\ref{CMB} shows the
comparison between all the models and the Planck CMB data. It is clear that the
CMB peaks of the Wetterich model are shifted to lower multipoles compared to $\Lambda$CDM.
All the other models have similar CMB spectra and fit the Planck data reasonably
well. At very low multipoles, $l<50$, the ``$\Lambda$CDM $1\sigma\ \Omega_{\rm m}$''
and ``$\Lambda$CDM $1\sigma\ H_0$'' models are almost the same as $\Lambda$CDM.
However, the two Doran \& Robbers models deviate from $\Lambda$CDM by up to 4 percent at
these multipoles. Hence the differences between the EDE1 and EDE2 models and
$\Lambda$CDM are not due to the fact that the EDE models are not formally the best
fitting models but rather arise because of the different expansion histories.
Fig.~\ref{CMB} also shows that the acoustic oscillations appear at slightly
different $l$ in EDE1 and EDE2 than in $\Lambda$CDM, as shown by the oscillations
in the ratio of power spectra shown in the lower panel.

\begin{figure}
\includegraphics[clip,width=8.5cm]{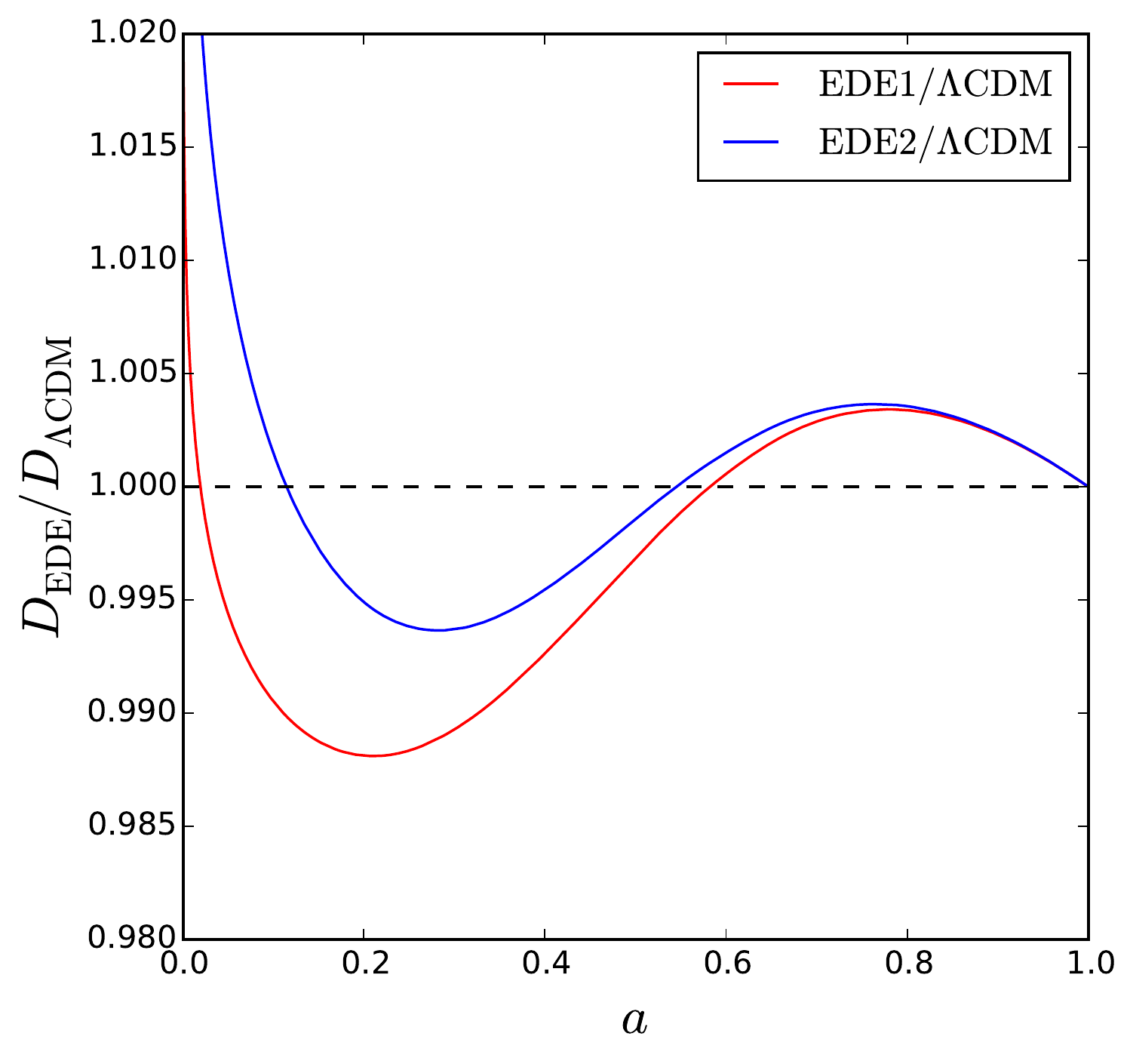}
\caption{The ratio of the linear growth factor in the EDE models considered here compared to $\Lambda$CDM as labelled. The linear growth factor is normalized to unity at $z=0$ in all models.}
\label{fig:GF}
\end{figure}

\begin{figure}
\includegraphics[clip,width=8.5cm]{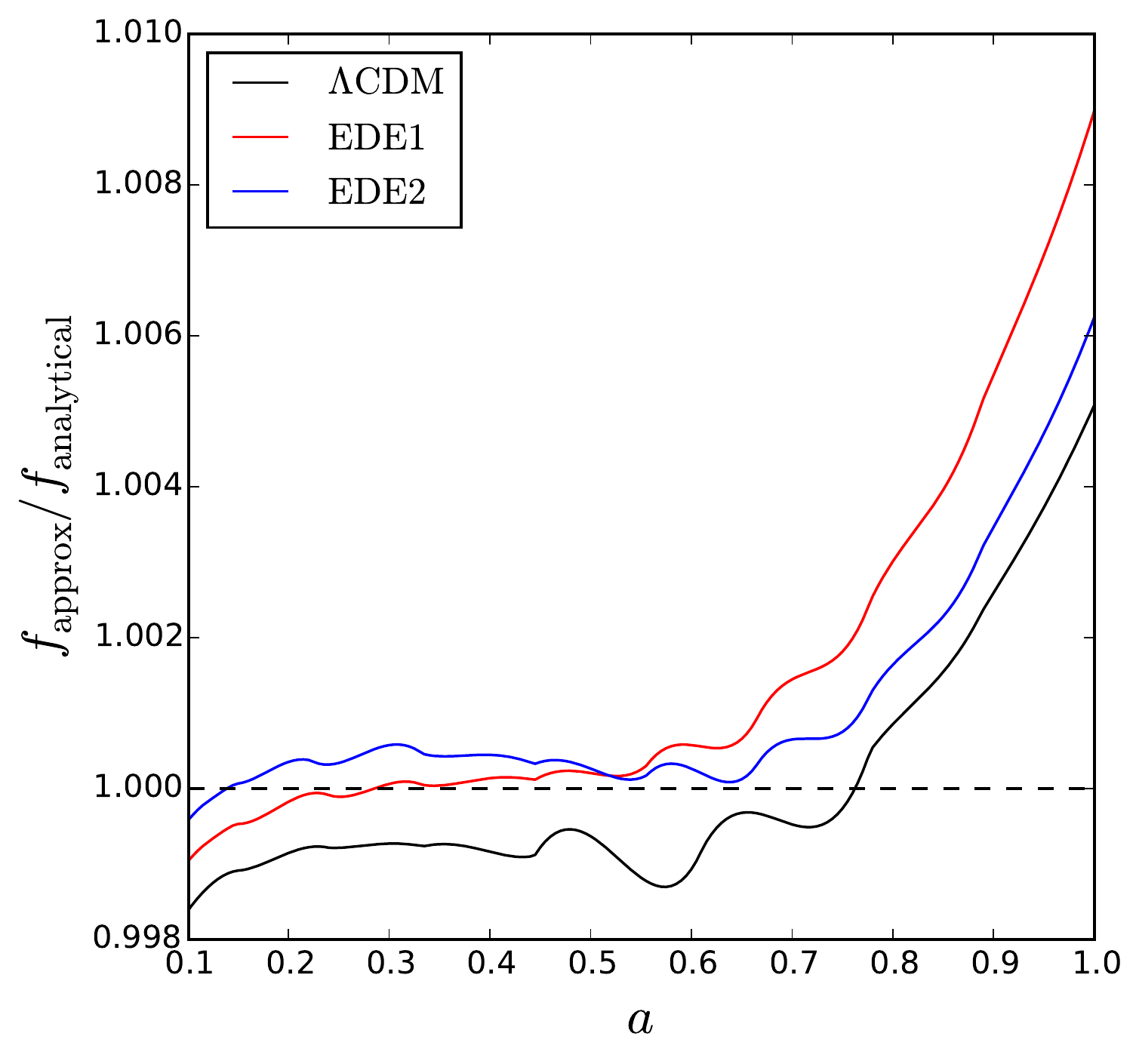}
\caption{A comparison of growth rate obtained using an approximation, $f_{\rm approx}$, estimated from Eqn.~\ref{eqn:f_approx} and the
analytical value, $f_{\rm analytical}$ calculated using Eqn.~\ref{eqn:GrowthFactor} in the EDE and $\Lambda$CDM models as labelled.}
\label{fig:GR_ratio}
\end{figure}

\subsection{Linear growth rate}
\label{subsec:GR}
The evolution of linear growth rate reflects the different growth histories of
structure between the EDE and $\Lambda$CDM cosmologies. If we assume the dark
matter perturbations are small, i.e., the density contrast $\ll 1$, the power
spectrum, $P(k,t)$ can be written as a function of time,
\begin{equation}
P(k,t)=\frac{D(t)^2}{D(t_0)^2}P(k,t_0).
\label{}
\end{equation}
Here, $D(t_0)$ is the linear growth factor today and is obtained by solving the
differential equation \citep{Linder1998}:
\begin{equation}
D^{\prime\prime}+\frac{2}{3}\left(1-\frac{w(a)}{1+X(a)}\right)\frac{D^{\prime}}{a}-\frac{3}{2}\frac{X(a)}{1+X(a)}\frac{D}{a^{2}}=0,
\label{eqn:GrowthFactor}
\end{equation}
where
\begin{equation}
X(a)=\frac{\Omega_{\rm m}}{1-\Omega_{\rm m}} e^{-3\int_{a}^{1} d \ln a^{\prime}w(a^{\prime})}.
\label{eqn:f_analitical}
\end{equation}
The linear growth rate is defined as $f=d\ln D/d\ln a$.
Fig.~\ref{fig:GF} shows the ratio of the linear growth factor in the EDE1 and EDE2
models to that in $\Lambda$CDM. Before $z=10$, the growth factor is enhanced by a few
percent in the EDE1 and EDE2 compared with $\Lambda$CDM, before showing a reduction for
$z \sim 2$--$10$.

Although it is straightforward to obtain the linear growth factor by solving
Eqn.~\ref{eqn:GrowthFactor}, some parametrizations of linear growth rate have
become popular. \citet{Peebles1976} proposed a widely used parametrization,
$f(z)\approx \Omega_m^{\gamma}$, where $\gamma=0.6$ is the growth index.
\citet{Linder2005} suggested the more accurate form
\begin{equation}
\label{eqn:f_approx}
\gamma=0.55+0.05[1+w(z=1)],
\end{equation}
which gives $f=\Omega_m^{0.55}$ for a $\Lambda$CDM cosmology.

In order test the accuracy of the Linder parametrization for the growth factor,
we plot  in Fig~\ref{fig:GR_ratio} the approximate growth rate, $f_{\rm approx}$, given by
Eqn.~\ref{eqn:f_approx} divided by the value $f_{\rm analytical}$ calculated
from Eqn.~\ref{eqn:GrowthFactor}. For the EDE1 and EDE2
models and $\Lambda$CDM, the approximation reproduces the linear growth rate
to better than 1\% over the redshift range from $z=0$ up to $z=10$. Nevertheless,
at late times the inaccuracy in the growth rate obtained from Eqn.~\ref{eqn:f_approx}
is comparable to the magnitude of the departure from the $\Lambda$CDM growth rate,
which means that the full calculation should be used.

\begin{figure}
 \includegraphics[width=8.5cm]{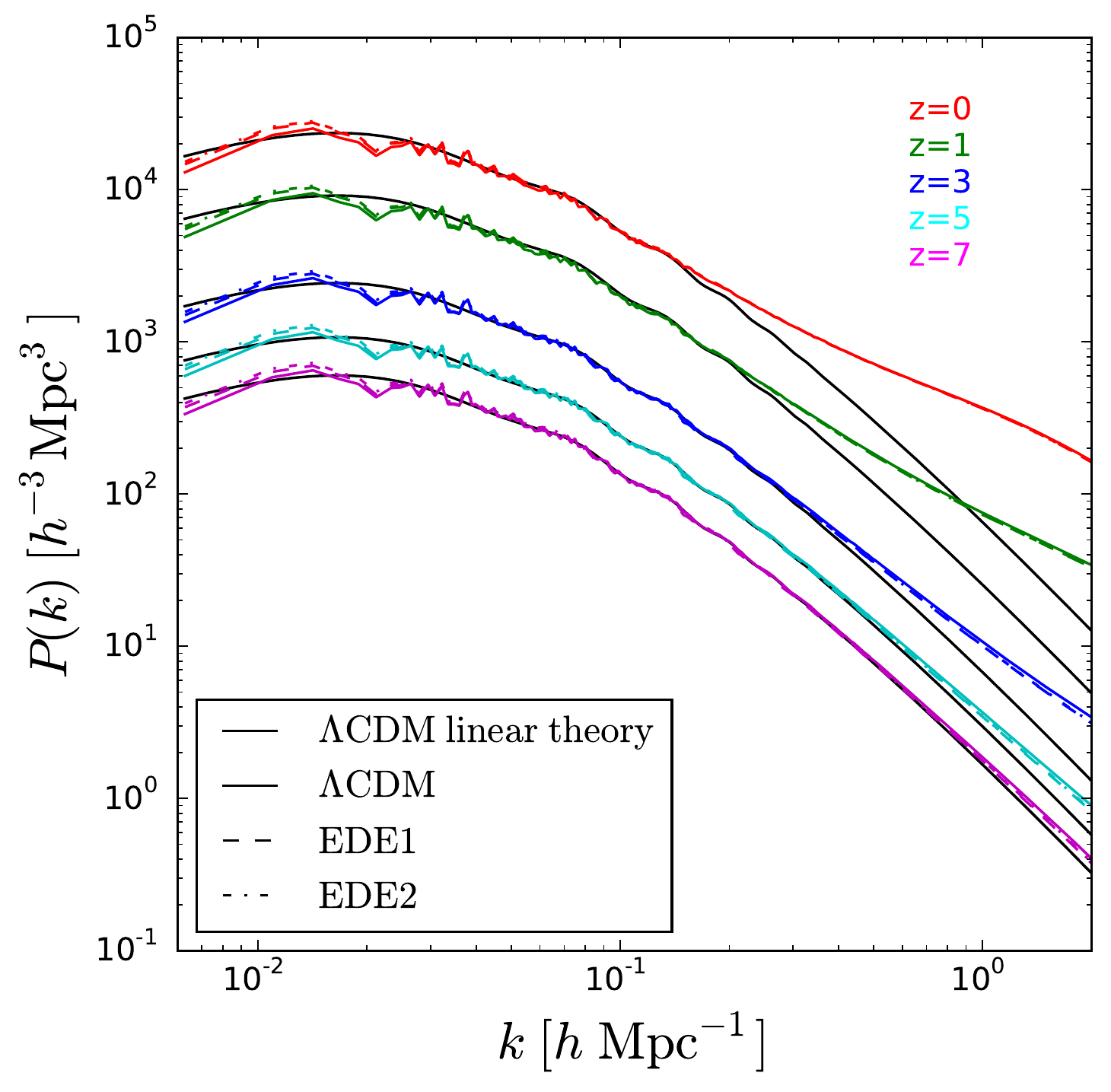}
 \caption{The matter power spectra measured in the EDE1, EDE2 and
$\Lambda$CDM simulations. Different line styles refer to the
results for different models and different colours show the
measurements at different redshifts, as indicated by the key.
The smooth black curves show the predictions of linear perturbation theory
in $\Lambda$CDM. Differences between the EDE1 and EDE2 models
and the $\Lambda$CDM results are apparent at very small
and high wavenumbers.
 }
 \label{fig:PowerSpec}
\end{figure}

\subsection{N-body Simulations}
\label{sec:simulation}

We have carried out three large volume, moderate resolution N-body simulations
for $\Lambda$CDM and the EDE1 and EDE2 cosmologies, using a memory-efficient
version of the TreePM code {\scshape GADGET-2}\citep{Springel2005}, called {\scshape L-GADGET2}.
The code was used in \citet{Jennings2010} and has been modified in order
to allow a time dependent equation of state for dark energy. We assume a
flat universe and use the cosmological parameters in Table~\ref{tab:best_fit}.
The simulations use grid initial conditions with $N=2048^3$ dark matter
particles in a computational box of a comoving length of $1500\ h^{-1}{\rm Mpc}$.
The particle mass is $3.413\times 10^{10} h^{-1}M_{\odot}$ for $\Lambda$CDM and
$3.064\times 10^{10} h^{-1}M_{\odot}$ for the EDE1 and EDE2 models. The initial
mean inter-particle separation is $0.732\ h^{-1}{\rm Mpc}$. We adopt a comoving softening
length of $\epsilon = 15\ h^{-1}\rm kpc$. The initial conditions were generated
using the {\scshape L-GENIC} code \citep{Springel2005b}, which
has also been adapted to handle a time variable equation of state.
A self-consistent linear theory power spectrum for each model is generated
using {\scshape CAMB} \citep{Lewis2002}. The normalisation extrapolated to $z=0$ is
$\sigma_8=0.8$ for all simulations. The starting redshift is $z=199$. We have
tested that the results presented have converged for these choices of particle
number, softening length and starting redshift. Here we are interested in
large-scale structure, redshift space distortions and rare objects, which is
why we chose a large simulation box.

\section{Results}
\label{sec:result}
Here we present a range of results from our N-body simulations:
the matter power spectrum in real and redshift space (\S~\ref{subsec:PS}),
the dark matter halo mass function (\S~\ref{subsec:HMF}),
and the distribution of counts-in-cells (\S~\ref{subsec:CIC}).

\subsection{Matter power spectrum}
\label{subsec:PS}

The power spectrum of fluctuations in the matter distribution is a key statistic that
encodes information about the cosmological parameters and is the starting point for
determining many quantities, such as the clustering of galaxies and the weak
gravitational lensing of faint galaxies. The presence of dark energy at early times
in the EDE cosmologies can change the form of the matter power spectrum compared
to that in $\Lambda$CDM and may allow us to distinguish between models.
The use of N-body simulations allows this comparison to be extended into the
nonlinear regime.

\subsubsection{The power spectrum in real space}

Fig.~\ref{fig:PowerSpec} shows the matter power spectra at redshifts
$z=0,1,3,5,7$ measured from the $\Lambda$CDM, EDE1 and EDE2 simulations,
together with the linear perturbation theory power spectra for $\Lambda$CDM.
At $z=0$, the power spectra have very similar amplitudes at intermediate
wavenumbers because the three models have been normalized to have the same
value of $\sigma_{8}$ today ($\sigma_8=0.8$). The power spectra, however,
are noticeably different at very small wavenumbers (large scales). There are
also small differences apparent deep into the nonlinear regime at high wavenumbers
(small scales).

\begin{figure}
 \includegraphics[width=8.5cm]{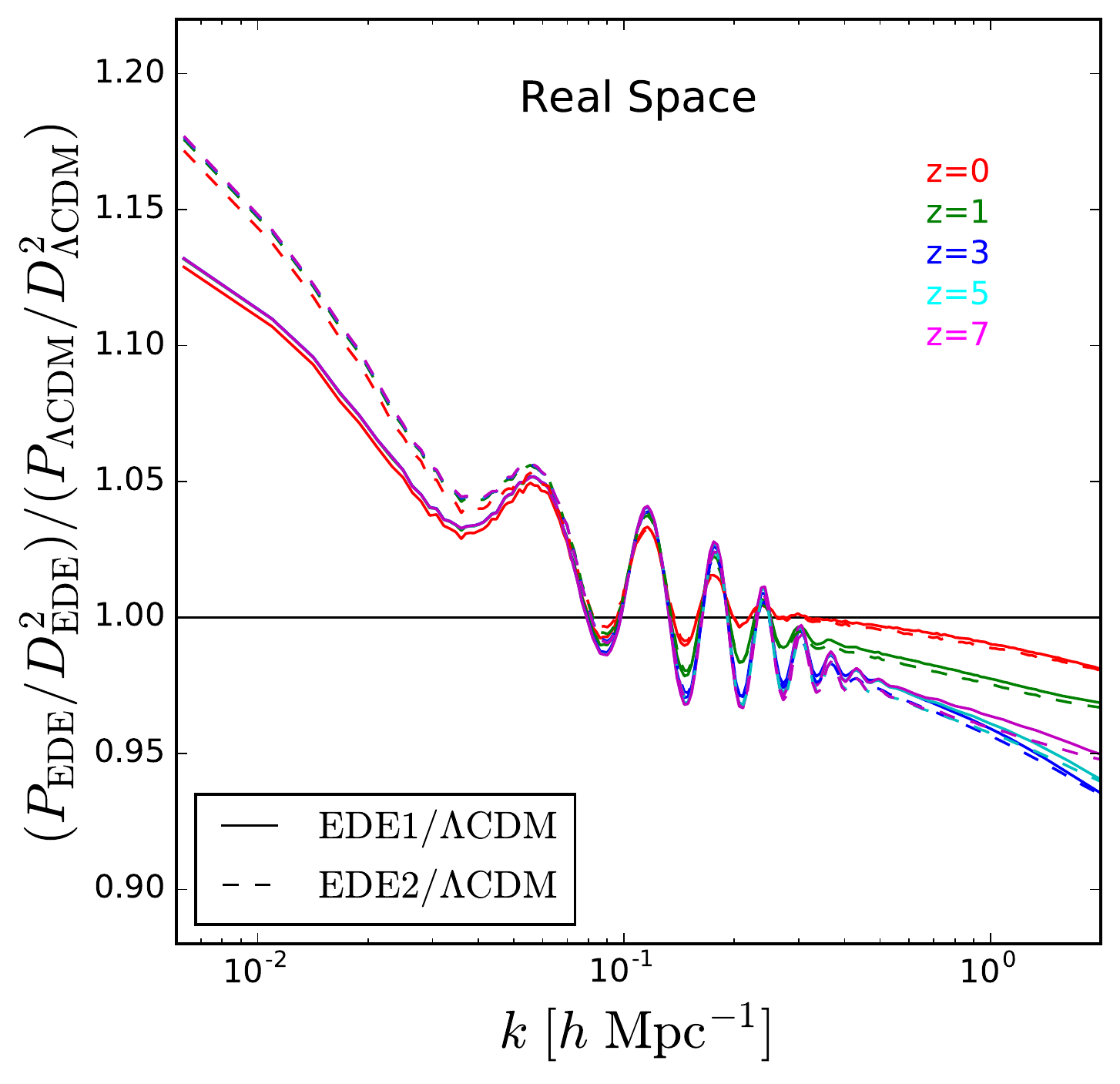}
 \caption{The ratio of matter power spectra measured in
real space in the EDE1 and EDE2 simulations to those in $\Lambda$CDM.
This ratio is plotted after taking into account differences in the linear growth factor at a fixed redshift between the models. The
differences on large scales (small $k$) show that it is important to
use a linear theory power spectrum in the simulations that is consistent
with the expansion history and cosmological parameters in the EDE models.}
 \label{PowerSpec_ratio}
\end{figure}


The EDE models differ from $\Lambda$CDM on large scales at all
plotted redshifts. This is due to the difference in the expansion histories
in these models compared with that in $\Lambda$CDM. This changes
the rate at which fluctuations grow, particularly around the transition
from radiation to matter domination, which alters the shape of the
turnover in the power spectrum \citep{Jennings2010}.
To drill down further into the comparison between the power spectra in the
models we now compare the simulation measurements after taking into account
differences in the linear growth factor at a given redshift (as plotted in
Fig.~\ref{fig:GF}). Fig.~\ref{PowerSpec_ratio} shows the
ratio of matter power spectrum after dividing by the linear growth
factor squared, $D(a)^2$, for each model.  The EDE1 and EDE2 models differ from $\Lambda$CDM by up to 13\% and 17\% on
large scales repectively, with the ratio showing a slight dependence on redshift. But the differences between the
models on small scales (high $k$) are more modest, reaching at most around 5\%.
Using the linear growth factor in
this way helps to isolate the impact of the different expansion histories in
the models (see \citealt{Jennings2010} for a more extended discussion of this comparison).
When plotted in this way, the ratios of power spectra measured at different redshifts
coincide. The residual differences at high wavenumbers are due to
the different growth histories in the models.

The non-negligible difference in Fig.~\ref{PowerSpec_ratio} illustrates the need to use a consistent
linear theory power spectrum to generate the initial conditions in
the N-body simulation rather than using a $\Lambda$CDM spectrum in all cases.

The conclusion of this subsection is that it should be possible to distinguish
an EDE model from $\Lambda$CDM using the shape of power spectrum on large scales, well
into the linear perturbation theory regime. The bulk of observational measurements
of the power spectrum probe the clustering in "redshift" space, so next we extend the
comparison to include the contribution from gravitationally induced peculiar velocities.

\subsubsection{The power spectrum in redshift space}

\begin{figure}
 \includegraphics[width=8.5cm]{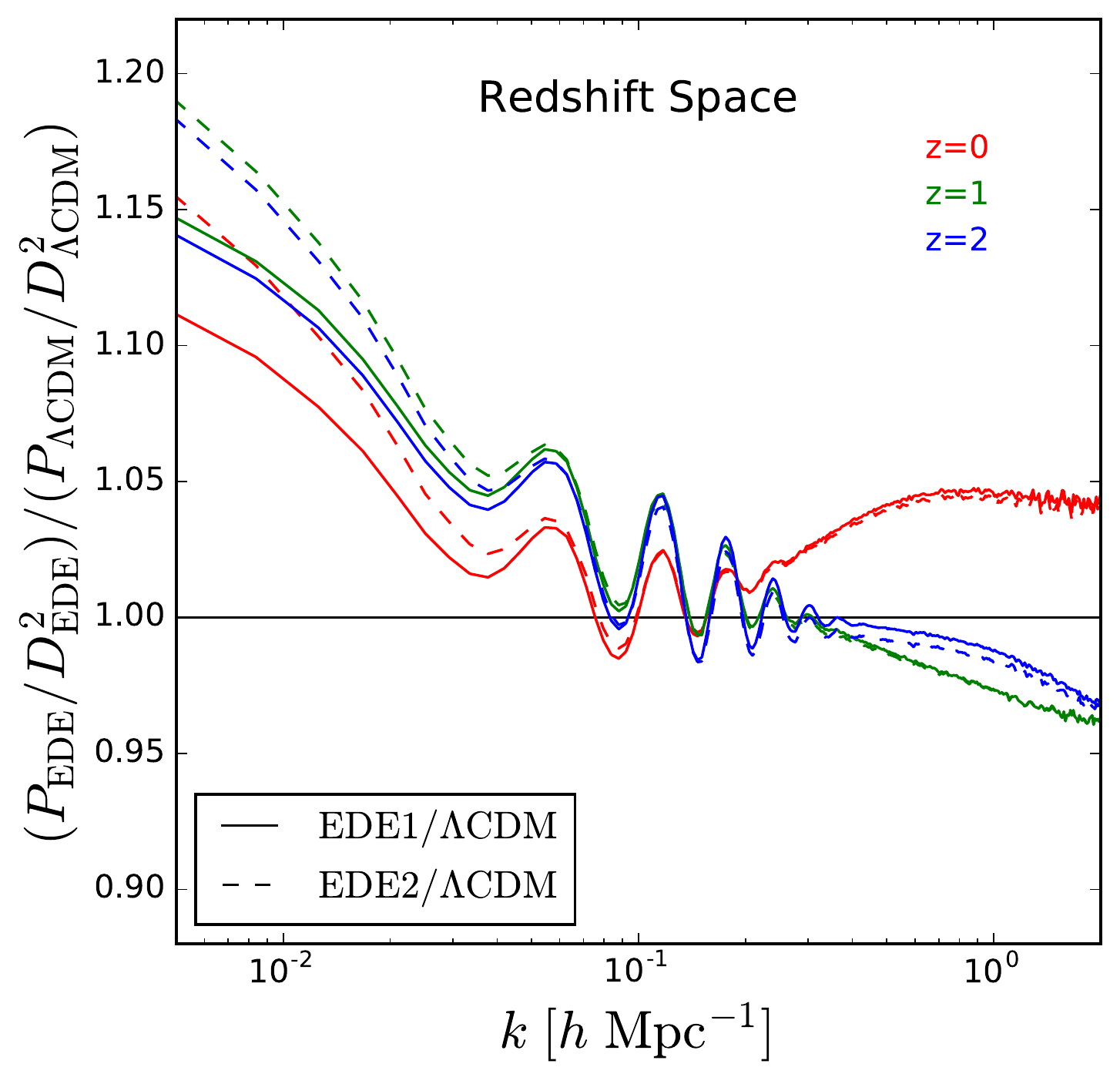}
 \caption{The ratio of redshift space power spectra
measured in the EDE1 and EDE2 simulations after dividing by the square of the linear growth
factor in each model  at the redshift in question to that in $\Lambda$CDM as labelled (note that the range of redshifts compared in
this plot is smaller than in Fig.~\ref{PowerSpec_ratio}).}
 \label{zPowerSpec_ratio}
\end{figure}

We model clustering in redshift space using the distant observer approximation.
We adopt one axis as the line of sight direction and displace the particles along
this axis according to the component of their gravitationally induced
peculiar velocity in this direction. Even though we use a large simulation volume,
there is still appreciable scatter in the clustering when viewed in redshift space,
so we repeat this procedure for each axis in turn and average the results to obtain
our estimate of the matter power spectrum in redshift space.

Fig.~\ref{zPowerSpec_ratio} shows the ratio of the redshift space power spectra measured in the simulations after removing differences in the linear
growth factors of the Doran \& Robbers cosmologies to that in $\Lambda$CDM.
On large scales, the EDE power spectra are 10\% - 20\% higher in amplitude
than the $\Lambda$CDM power spectrum, which is similar to the result found in real space.
On small scales, due to the nonlinear effects, there are clear differences in the $P(k)$,
but these are smaller than 5 per cent.
However, unlike the case of the real space power spectra, dividing by growth
factor squared does not reduce the differences between the ratios measured at
different redshifts. Instead, the difference between the rations measured between the
redshift space power spectra in a given pair of models increases slightly on large
scales. This is because the linear growth factor does not account for all of the
linear theory differences between the power spectra in redshift space.

\begin{figure*}
\includegraphics[width=16cm]{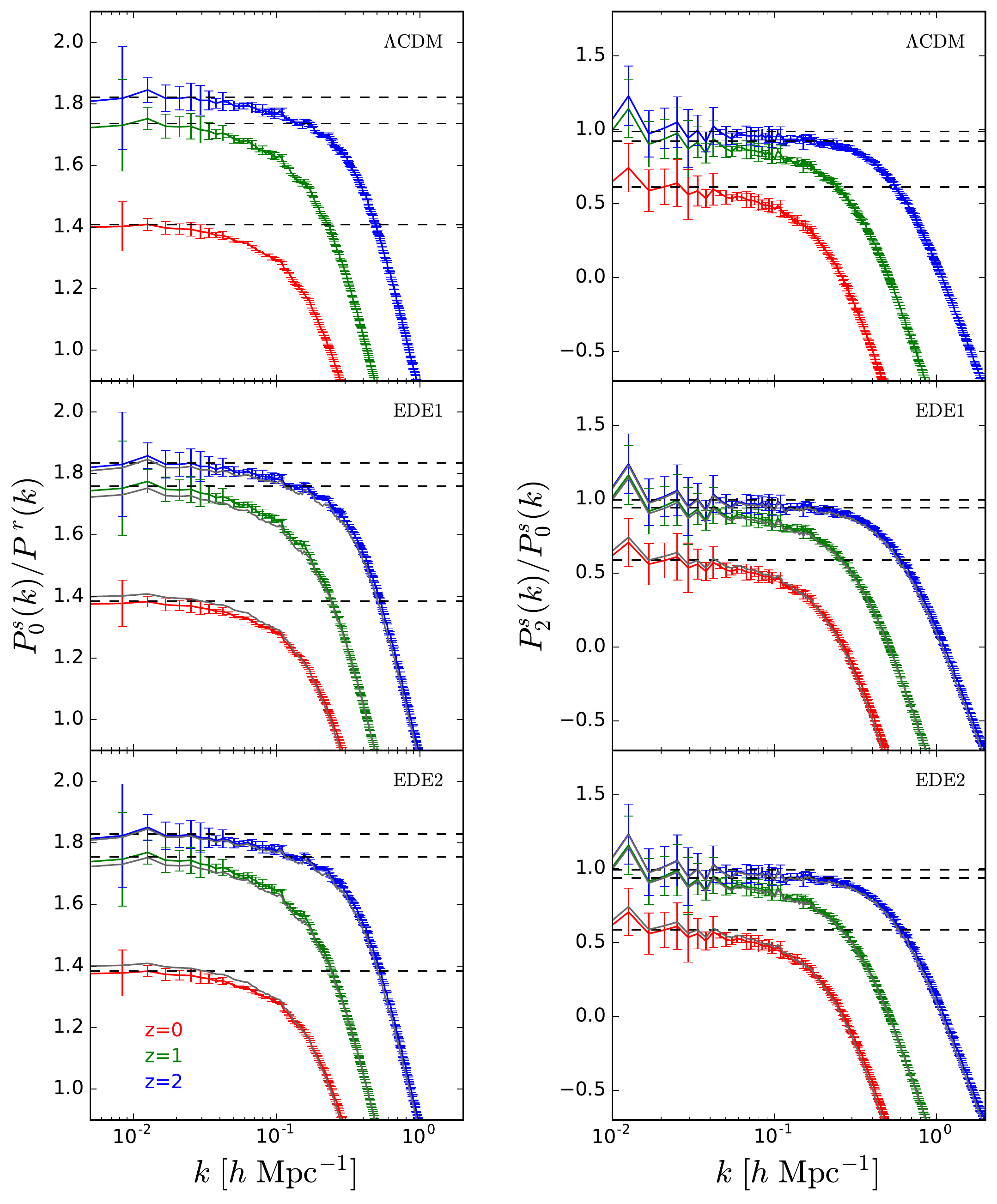}
\caption{The distortion of clustering due to peculiar velocities.
Left panel: the ratio of the monopole redshift power spectra to real
space power spectra measured from the N-body simulations at $z=0,1$ and $2$.
Different colours show the results for different redshifts as labelled.
The dashed lines show the linear theory prediction.
Right panel: the ratio of the quadrupole to monopole moments of the
redshift power spectra measured from the simulations. Each panel shows
the result for a different model as labeled. For comparison,
the $\Lambda$CDM measurements are reproduced as grey lines in the EDE1 and EDE2
panels}
\label{kaiser_ratio}
\end{figure*}

To further investigate the contributions of the velocity dispersion and nonlinearities to
the form of the redshift space power spectrum, we compare the ratio of the spherically
averaged power spectra in redshift space and real space in left column of
Fig.~\ref{kaiser_ratio}. The linear theory prediction, known as the ``Kaiser formula''
given by $P^{s}(k,\mu)=P^{r}(k)(1+\mu^2\beta)^2$, is plotted as black dashed lines
in Fig.~\ref{kaiser_ratio}. Here $P_{r} (k)$ is the power spectrum in real-space,
$\mu$ is the cosine of the angle between the line of sight and the peculiar motion of
the dark matter particle and $\beta = f$ for the dark matter.
The linear theory monopole ratio depends on redshift through the value of the matter density parameter.
The value of linear growth rate is calculated using the parametrization
$f(z)=\Omega^\gamma$, where $\gamma$ is given by Eqn.~\ref{eqn:f_approx}. The error bars
illustrate the scatter in $P^s(k)$ obtained by using the x, y, z directions in turn
as the line-of-sight direction.
At $z=0$, the left panel of Fig.~\ref{kaiser_ratio} shows that the Kaiser formula
only fits the simulation results on very large scales, $k<0.03 h \rm{Mpc^{-1}}$,
as reported by \citet{Jennings2011}. The departure from the linear theory prediction
is due to a combination of nonlinearities and the damping effects of peculiar velocities,
even though this is often modelled as arising solely due to damping.
Nonlinear effects are important for $k>0.03 h \rm{Mpc^{-1}}$ even though the
linear regime is typically believed to hold out to $k \sim 0.1$ -- $0.25 h \rm{Mpc^{-1}}$.
The Kaiser prediction agrees with the simulation results over a slightly wider
range of scales at higher redshifts because the nonlinear effects are smaller than
they are that at $z=0$. In the right panel of Fig.~\ref{kaiser_ratio} we plot the ratio
of the quadpole to monopole moments of the redshift space power spectrum,
$P^s_2(k)/P^s_0(k)$, for each cosmology at $z=0$, $1$ and $2$.
The Kaiser limit agrees with the simulation results for $k<0.05 h \rm{Mpc^{-1}}$
at $z=0$ which is a slightly higher wavenumber than was the case for the
monopole ratio. The departures from the redshift space distortions expected
in $\Lambda$CDM (shown by the grey lines in Fig.~\ref{kaiser_ratio}) are small,
and well within our estimated errors.

\begin{figure}
 \includegraphics[clip,width=8.5cm]{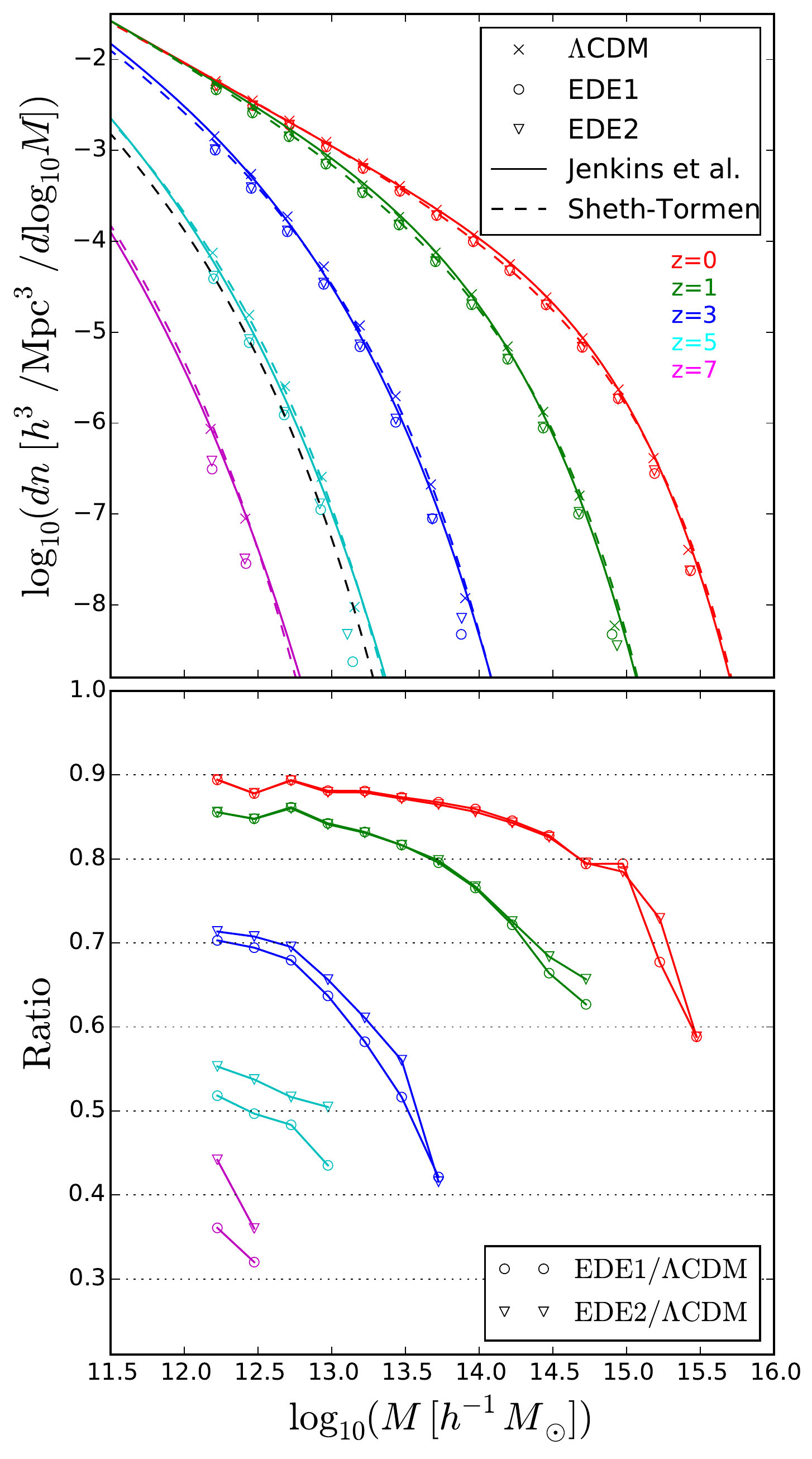}
 \caption{The mass function of dark matter halos measured from the simulations.
The upper panel shows the halo mass functions at different redshifts.
The crosses show $\Lambda$CDM, circles EDE1 and triangles EDE2. The solid and
dashed lines show the Jenkins et al. and Sheth-Tormen mass functions respectively for
$\Lambda$CDM. The lower panel shows the ratio of dark matter halo mass functions
in the EDE1 and EDE2 cosmologies to that measured in $\Lambda$CDM.}
 \label{fig:hmf}
\end{figure}

\subsection{Halo mass function}
\label{subsec:HMF}

The mass function of dark matter halos, defined as the number of halos per unit volume
with masses in the range $M$ to $M + dM$, $n(M, z)$, is an important characteristic
of the dark matter density field which is affected by the expansion rate of the Universe.

We use the friends-of-friends (FOF) algorithm \citep{Davis1985} which is built into
the L-GADGET2 code to identify dark matter halos, using a linking length of $b=0.2$ times the
mean inter-particle separation. We retain FOF groups down to 20 particles.
In Fig.~\ref{fig:hmf} we plot the halo mass functions measured from the $\Lambda$CDM,
EDE1 and EDE2 simulations at $z=0,1,3,7$. For comparison we also plot
the \cite{Jenkins2001} and \cite{Sheth1999} mass functions evaluated for $\Lambda$CDM.
The lower panel of Fig.~\ref{fig:hmf} shows the ratio of the mass functions measured
in the EDE cosmologies to that in $\Lambda$CDM.
The differences in the mass functions at low redshift ($ z \le 1$) are small, in
agreement with results of \citet*{Francis2009}. The EDE mass functions agree
with $\Lambda$CDM to within 20\% for halos with masses around $10^{12.0}$--$10^{13.5}
h^{-1}M_{\odot}$ at $z=1$.

The difference between the halo mass functions in EDE and $\Lambda$CDM increases
with increasing redshift. This is due in part to the difference in the linear
growth factors getting larger between the EDE and $\Lambda$CDM cosmologies going
back in time from the present day. Also, because the simulations have a fixed
mass resolution, the results probe rarer halos with increasing redshift. The abundance
of these objects is sensitive to the matter power spectrum at smaller wavenumbers,
where we found the largest differences between EDE and $\Lambda$CDM.
At $z=7$, $\Lambda$CDM predicts 2.5 times as many halos as are found in the
EDE cosmologies.

\begin{figure*}
\includegraphics[clip,width=15cm]{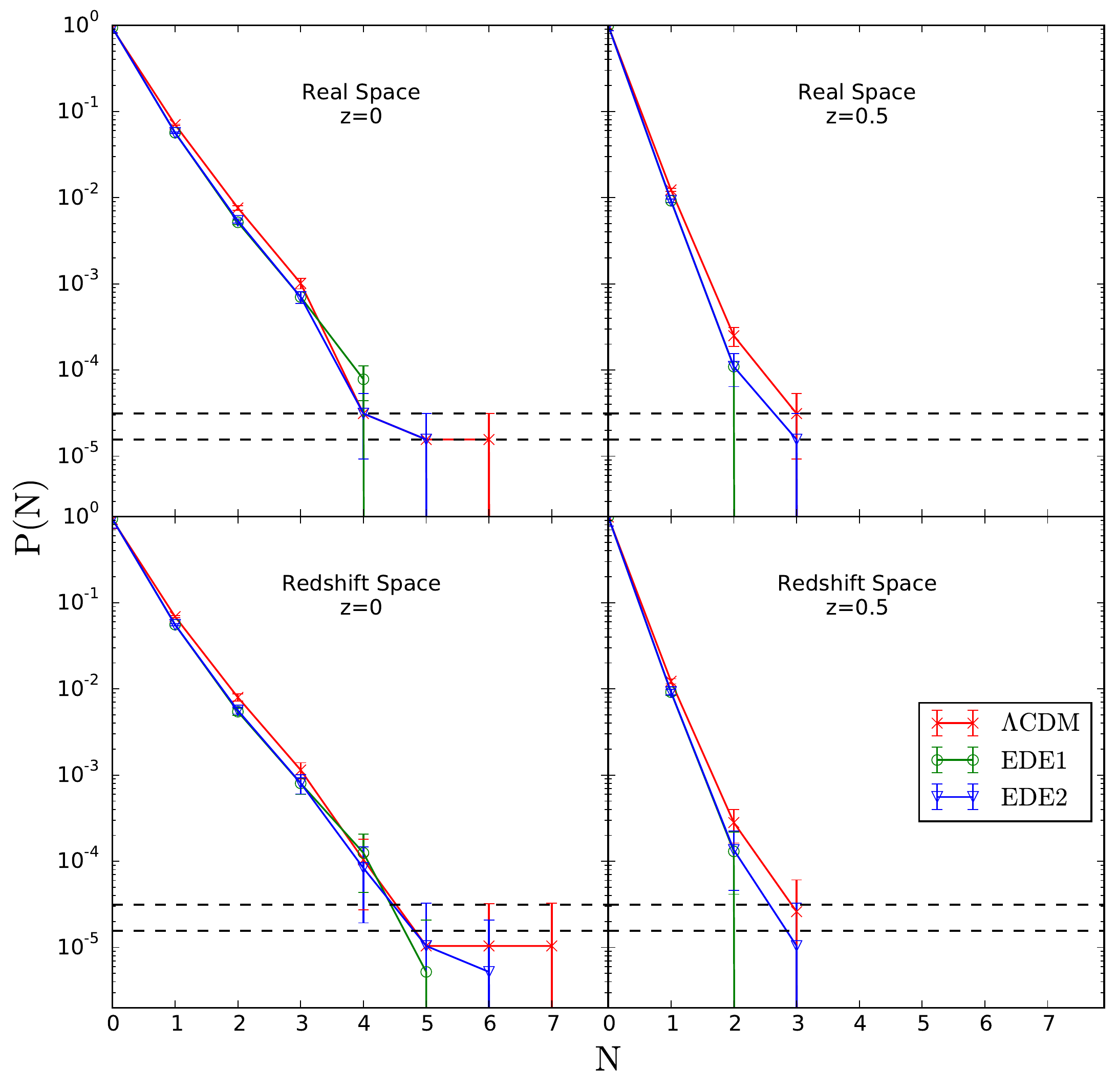}
\caption{ The counts in cells distribution for extreme structures.
The probability distribution of finding a given number of halos with
mass larger than $5\times10^{14}\ h^{-1}\rm M_\odot$ in cubic cells
of side $37.5\ h^{-1}\rm Mpc$. The x-axis is the number of halos in the
cell. In the upper panels, the cell counts are measured in the real space, while
in the lower panels the counts are measured in redshift space.
The two dashed horizontal lines in each panel indicate the probability to
find one cell and two cells in the whole simulation box.}
\label{fig:massive_halo}
\end{figure*}

This prediction could be tested by using a proxy for the halo
mass function at high redshift, such as the galaxy luminosity
function (\citet{Jose2011} proposed a similar test to probe
the mass of neutrinos). To make the connection to the observable
Universe, a model is needed to connect the mass of a dark matter
halo to the properties of the galaxy it hosts. We have evaluated this
approach by carrying out an abundance matching exercise between
the halo mass functions and the observed luminosity function of
galaxies in the rest-frame ultra-violet. This simple procedure
assigns one galaxy to each dark matter halo, ignoring any contribution
from satellite galaxies. The translation between halo mass and galaxy
luminosity can be described by a mass-to-light ratio. Despite the
large differences in the halo mass functions between cosmologies,
the differences in the implied mass-to-light ratios are quite
modest and well within the current uncertainties in our knowledge
of the galaxy formation process. Hence, we conclude that any of these
cosmologies could be made to match the observed galaxy luminosity function
at high redshift with plausible mass to light ratios, and that it would be
difficult to use the galaxy luminosity function to distinguish between
the models.

\subsection{Extreme structures}
\label{subsec:CIC}

We have seen in Section~\ref{subsec:PS} that the power spectra of the $\Lambda$CDM and
EDE energy models are similar on small scales, particularly once the differences
between the expansion histories in the models have been taken into account. The
power spectrum is a second moment of the density field and so does not probe the
tails of the distribution of density fluctuations, which could carry the imprint of differences
in the growth history of fluctuations.

Fluctuations in the density field can be quantified by measuring the distribution of
fluctuations smoothed over cells, commonly referred to as counts-in-cells.
Rather than formally measuring the higher order moments of the counts-in-cells
distribution, which rapidly becomes infeasible even with simulations of the volume
used here, we instead compare the high fluctuation tails of the distributions directly
in different cosmologies.

Following \cite*{Yaryura2011}, in order to connect more closely with observables
rather than looking at fluctuations in the overall matter distribution, we consider
the counts-in-cells of cluster-mass dark matter halos. In particular, we look
for ``hot" cells that contain a substantial number of massive halos.
The choice of halo mass and the definition of hot cells is motivated by
results from the two-degree field galaxy redshift survey (2dFGRS).
\citet{Croton2004} identified two hot cells in the 2dFGRS. \citet{Padilla2004}
found 10 groups with an estimated mass over $5\times10^{14}\ h^{-1}\rm M_\odot$
in each cell, by cross matching the hot cells in the galaxy distribution
with the 2dFGRS Percolation Inferred Galaxy Group catalogue \citep[2PIGG catalogue,][]{Eke2004a}.

Here we use a cubical cell of side $37.5\ h^{-1}\rm Mpc$, which corresponds to a
slightly smaller volume than the equivalent size of the spherical cell used by \citet{Croton2004}.
We then count the number of dark matter halos with $M_{\rm halo}>5\times10^{14}\ h^{-1}\rm M_\odot$
inside each cell. We use the jackknife method to estimate the errors on the distribution of counts \citep{shao1986,Norberg2009} .
Put simply, the jackknife is a resampling technique which works by systematically leaving out each subset of data in
turn from a whole dataset to generate ``new'' subsamples. Here a subset is defined to be a volume within the simulation.
Then, the overall jackknife estimate of $\delta$ can be found by averaging over all the subsamples, given by
\begin{equation}
\delta_{\rm Jack}=\frac{1}{N}\sum_{i=1}^N \delta_i,
\label{}
\end{equation}
where $N$ is the number of subsamples. The jackknife error is calculated as
\begin{equation}
\sigma_{\rm Jack}=\sqrt{(N-1)\sum_{i=1}^N \frac{(\delta_i-\delta_{\rm Jack})^2}{N}}.
\label{}
\end{equation}
We use 64 spatial subsamples in our analysis, dividing each side of simulation box equally into four parts.

Fig.~\ref{fig:massive_halo} shows the distribution of cell counts for the three cosmologies in both real space
and redshift space at $z=0$ and $0.5$. The $x$-axis gives the number of halos per cell above the specified mass limit.
The $y$-axis is the normalized probability to find such a cell.
In redshift space, we also considered the scatter from using the three axes in turn as the line-of-sight.
The high cell count tails are very similar, but $\Lambda$CDM consistently predicts more ``hot'' cells.
The ``hottest'' cells only contain 7 halos in $\Lambda$CDM at $z=0$, which is lower than suggested by the 2dFGRS superclusters.
This could be because the FOF halo mass, which we used to select the halos, does not match the halo mass estimated from the
galaxy group catalog. \citet*{Yaryura2011} showed that by perturbing the FOF halo mass by the systematic bias and scatter
expected in the masses returned by a group finder run on a galaxy catalogue, the number of hot cells increases.

Again, the differences between the predicted count distributions are smaller than the estimated errors
on the measurement and so could only be probed by a survey covering a volume that is much larger than our simulations.

\section{Conclusions}
\label{sec:conclusion}

One of the main science goals of future wide field galaxy surveys is
to distinguish a cosmological constant from other scenarios for the
acceleration of the cosmic expansion, such as dynamical dark energy models.
Here we have examined a particular class of dynamical dark energy model which
display a small but non-negligible amount of dark energy at early times,
which are referred to as early dark energy models. Such models could be motivated by a
choice of potential for the scalar field describing the dark energy. Instead,
to confront these models with the currently available cosmological constraints
in an efficient way, we chose to use a simple description
in which the density parameter of the dark energy is parametrized as a function
of the expansion factor, the present day values of the dark energy and matter
density parameters, the present equation of state parameter of the dark energy
and the asymptotic value of the density parameter of dark energy at early times.
Once constrained, the model can be described by the resulting time dependence of
the equation of state parameter.

The step of constraining the early dark energy model to reproduce current
observations is a critical one. In fact, the best fitting models, even with the
observational constraints available today favour models {\it without} any dark
energy at early times, a conclusion that has already been reached by other studies
\citep{Planck2013XVI}. Nevertheless, within the range of models that remain
compatible with current data, it is possible to find examples with interesting amounts
of early dark energy. Increasing the amount of early dark energy in the model tends
to favour a more negative equation of state parameter at the present day than the
canonical $w=-1$ which corresponds to the cosmological constant. We have investigated
two models which, whilst not best fitting models, are still compatible with the
observations at the $1-\sigma$ (EDE1 with 1\% of the critical density in dark energy
at early times) and $2-\sigma$ levels (EDE2 with 2\% of the critical density in dark energy at
early times); both models have $w_{0} = -1.2$.

Previous simulation work on early dark energy models suggested that a clear signature
that could be testable against $\Lambda$CDM is the halo mass function
\citep*{Francis2009, Grossi2009}. In a simple picture, the presence of a small but
unignorable amount of dark energy at early epochs increases the rate at which the
universe expands, making it harder for structure to form. If the models are set up to
have the same value of $\sigma_{8}$ today, this means that structure has to form at a
smaller expansion factor or earlier time in the early dark energy model. Hence, a larger
number density of massive haloes is predicted in early dark energy models compared to
$\Lambda$CDM.

Our results show that this simple picture of early structure formation with early
dark matter is not a generic feature of these models. After constraining the models
against current observations, we find that the evolution of the linear
growth rate of fluctuations in the early dark energy models is remarkably close
to that in $\Lambda$CDM. At the earliest epochs, the EDE2 growth rate exceeds that
in $\Lambda$CDM by just $2\%$ before lagging behind until catching up around $z \sim 0.8$
and then exceeding the $\Lambda$CDM growth rate by less than $0.5\%$.

The dark matter halo mass function in the early dark energy simulations shows {\it fewer} massive
haloes than we find in the $\Lambda$CDM simulation. This difference in the halo mass function
could be tested using the high redshift galaxy luminosity function (as suggested by \cite{Jose2011}
to probe the nature of massive neutrinos). The difference in halo abundance is, however,
modest, and could be accounted for by our lack of knowledge of the relevant galaxy formation
physics. We find a small difference in the abundance of ``hot cells" in the distribution of
dark matter halos between early dark energy and $\Lambda$CDM, though this will be challenging to
measure, requiring huge survey volumes.

The cleanest signature we have found of the presence of dark energy at early times is
in the shape of the matter power spectrum. The more rapid expansion rate around the
epoch of matter radiation equality in early dark energy models compared to $\Lambda$CDM
changes the shape of the turn over in the matter power spectrum \citep{Jennings2010}. This
effect is visible in the linear theory power spectrum and is present on scales on which
we would expect scale dependent effects in galaxy bias to be small \citep{Angulo2008}.
To probe this effect it will necessary to retain the full shape information for the
galaxy power spectrum, rather than isolating the scale of the baryonic acoustic oscillation
feature \citep*{Sanchez2008}. \tcr{Tentative measurements of the matter power spectrum on the scale of the turnover
have already been made by the WiggleZ Dark Energy survey \citep{Poole2013}. Future large-area radio surveys conducted with the SKA pathfinder experiments, MeerKAT and ASKAP have the potential to probe the existence of early dark energy by providing
more accurate measurements of the turnover in the power spectrum. }

\section*{Acknowledgments}

DS acknowledges support from a European Research Council Starting Grant (DEGAS-259586) and Euclid implementation phase (ST/KP3305/1). CMB acknowledges receipt of a Research Fellowship from the Leverhulme Trust. This work was supported by the UK Science and Technology Facilities Council (STFC) Grant no.~ST/L00075X/1. This work used the DiRAC Data Centric System at Durham University, operated by the Institute for Computational Cosmology on behalf of the STFC DiRAC HPC Facility (http://www.dirac.ac.uk). This equipment was funded by BIS National E-infrastructure capital grant ST/K00042X/1, STFC capital grant ST/H008519/1, STFC DiRAC Operations grant ST/K003267/1 and Durham University. DiRAC is part of the National E-infrastructure. We thank the reviewer for providing a helpful report.

\label{lastpage}

\bibliographystyle{mn2e}
\bibliography{EDE}
\end{document}